\documentclass[twocolumn]{aastex631}

\begin{document}

\title{Discovery of metal-poor and distant pre-main sequence candidates in WLM with JWST}

\author[0000-0002-4641-2532]{Venu M. Kalari}
\affiliation{Gemini Observatory/NSF’s NOIRLab, Casilla 603, La Serena, Chile}
\author[0000-0002-1206-1930]{Ricardo Salinas}
\affiliation{Nicolaus Copernicus Astronomical Center, Polish Academy of Sciences, Bartycka 18, 00-716 Warszawa, Poland}
\author[0000-0002-5306-4089]{Morten Andersen}
\affiliation{European Southern Observatory, Karl-Schwarzschild-Strasse 2, 85748 Garching bei Mu{\"n}chen, Germany}
\author{Guido De Marchi}
\affiliation{European Space Agency (ESA), European Space Research and Technology Centre (ESTEC), Keplerlaan 1, 2201 AZ Noordwijk, The Netherlands}
\author[0000-0002-5307-5941]{Monica Rubio}
\affiliation{Departamento de Astronom\'ia, Universidad de Chile, Santiago, Chile}
\author[0000-0002-8445-4397]{Jorick S. Vink}
\affiliation{Armagh Observatory and Planetarium, Division or Department, College Hill, BT61 9DG Armagh, UK}
\author{Hans Zinnecker}
\affiliation{Universidad Aut\'onoma de Chile, Pedro de Valdivia 425, Providencia, Santiago de Chile, Chile}



\begin{abstract}

We present the discovery of twelve metal-poor and distant pre-main sequence (PMS) candidates in the dwarf irregular galaxy Wolf–Lundmark–Melotte (WLM) $\sim$968\,kpc away, at a present-day metallicity of [Fe/H]\,$\sim$\,$-$0.9. These candidates have masses between 1.25--5\,M$_{\odot}$, with ages $<$10\,Myr, and exhibit significant near-infrared excesses at 2.5 and 4.3\,$\mu$m. They are concentrated within a cluster roughly 10\,pc (2$\arcsec$) across situated in the H{\scriptsize II} region [HM95]-9. These are the most distant and metal-poor PMS stars known, and can offer new quantitative insights into star formation at low-metallicities.


\end{abstract}



\section{Introduction} \label{sec:intro}

In the current paradigm of star formation, cold molecular clouds collapse into protostars. Mass is then accreted until internal fusion is ignited. This picture is expected to differ quantitatively as a function of metallicity ($Z$). 

Reduced cooling via metal lines inhibits cloud fragmentation which may lead to higher average stellar masses and a top-heavy initial mass function at lower metallicities \citep{kroupa01}. This has been detected at $Z\sim$0.2--0.5\,$Z_{\odot}$ (\citealt{marks, schndr, ngc796, yasui2, zinkann}). Weaker stellar winds and feedback in metal-poor local group galaxies are also predicted to allow for more efficient star formation \citep{vink}. Observationally, low-$Z$ molecular clouds show low CO abundances, resulting in small, clumpy clusters (for e.g. \citealt{pak, schruba, inde, schruba2, archer} at 0.1--0.5\,$Z_{\odot}$). The lower dust-to-gas ratio permits UV radiation from massive stars to more effectively photo-evaporate circumstellar disks, reducing disk mass \citep{yasui, ercolano, guar, demarchi24} at sub-solar metallicities. These competing factors result in quantitatively different star formation efficiencies and histories as a function of metallicity. Studying these environments offers insights into the formation of stars and galaxies at cosmic noon ($z\sim$1; 0.1$Z_{\odot}$), when the bulk of star formation occurred \citep{lilly, madau}.

James Webb Space Telescope (JWST) can resolve individual stars in nearby dwarf galaxies (around $\sim$1\,Mpc) in the near to mid-infrared. JWST's deep infrared imaging is crucial for identifying young stars (under 10 Myr), which are often hidden in dusty envelopes or disks but reveal their presence through infrared excesses. This capability enables detailed studies of ongoing star formation in local group galaxies, in regions with conditions similar to those at cosmic noon.

In this letter, we utilize JWST early release science data \citep{weisz} of the dwarf galaxy WLM (Wolf–Lundmark–Melotte; or DDO\,221) to identify young pre-main sequence (PMS) candidates. In Section 2 we characterize the star-forming stellar population in WLM, while Section 3 focuses on the identification of PMS candidates. Finally, Section 4 outlines future work, and contextualizes our findings with respect to nearby star-forming regions.

\begin{figure}
\centering
\includegraphics[width=0.4\textwidth]{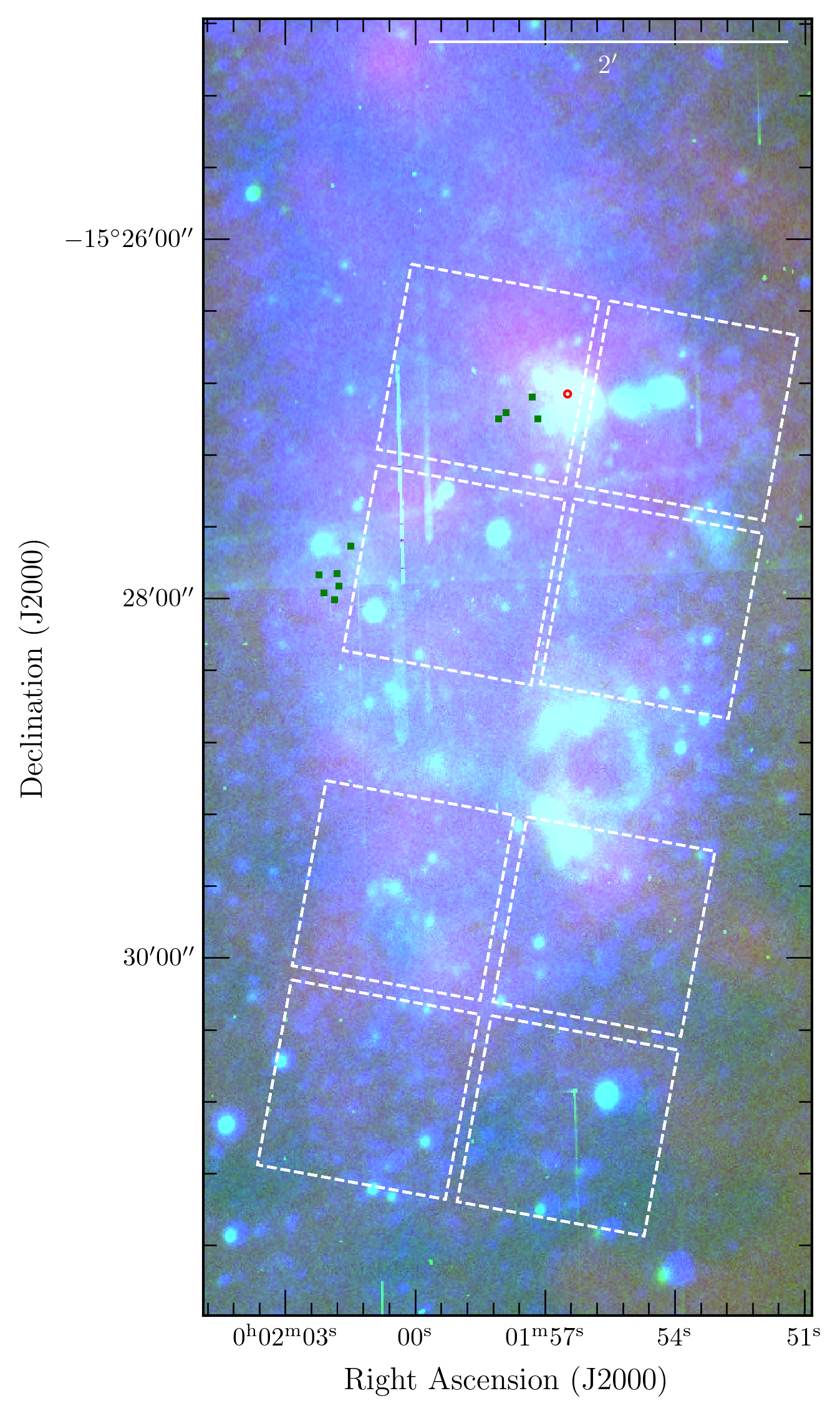}
\caption{WLM $r/g/b$ image where $r$=24\,$\mu$m, $g=$H$\alpha$, and $b$ is the optical $V$-band image. The red circle marks the cluster. North is up and east is to the left. Green squares mark the location of CO cores from \cite{rubio}. The dashed white lines represent the footprint covered by \cite{weisz}. }\label{whole}
\end{figure}

\begin{figure}
\centering
\includegraphics[width=0.45\textwidth]{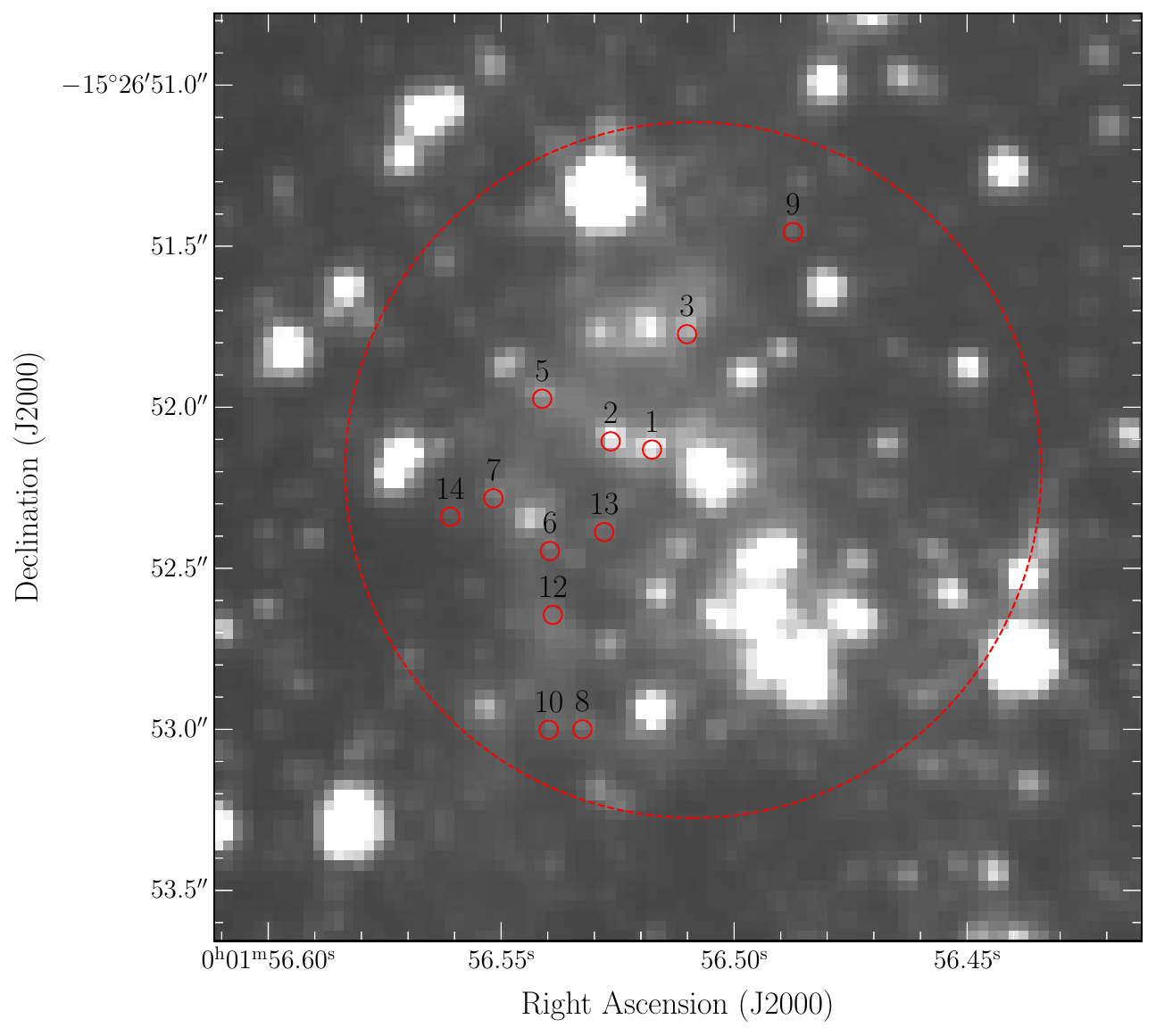}
\caption{F150W image of the cluster. The dashed red circle (radius of 1.1$\arcsec$) encircles the region studied here. PMS candidates are labeled following the identification given in Table 1.}\label{cl}
\end{figure}

\section{Star-forming stellar population in WLM} \label{sec:style}

\begin{figure}
\centering
\includegraphics[width=0.4\textwidth]{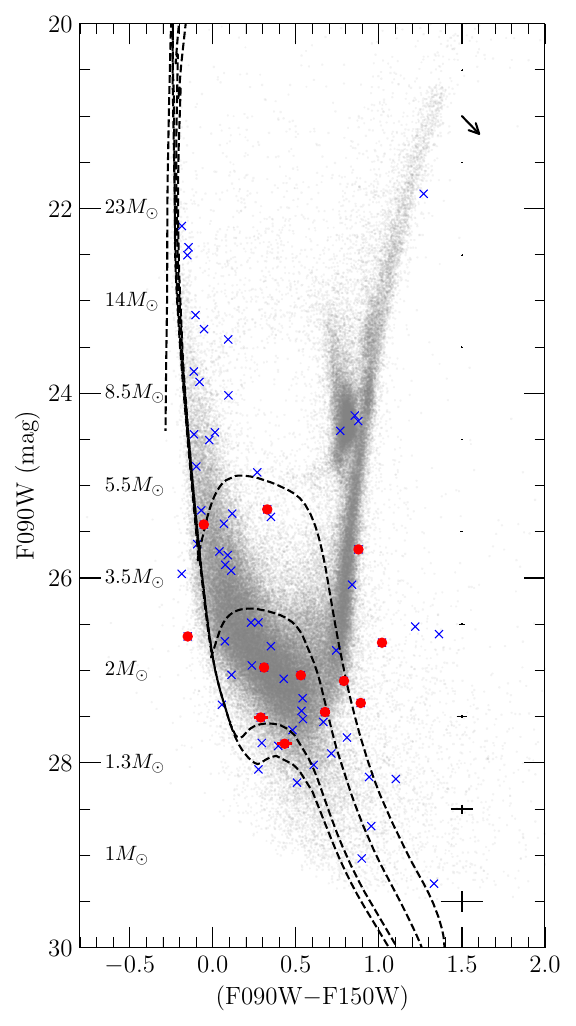}
 \caption{(F090W$-$F150W) vs. F090W color-magnitude diagram of the whole JWST FoV (gray dots), and the central cluster (blue crosses). Shown are the 1\,Myr, 3\,Myr, 8\,Myr, and 10\,Myr \cite{mist} isochrones. The stellar masses correspond to main-sequence stars at the distance and extinction adopted. The pre-main sequence candidates selected using the two-color diagrams are shown as red circles. The extinction vector corresponds to the adopted foreground and cluster extinction value (see text). }\label{cmd}
\end{figure}

\begin{figure*}
\centering
\includegraphics[width=0.45\textwidth]{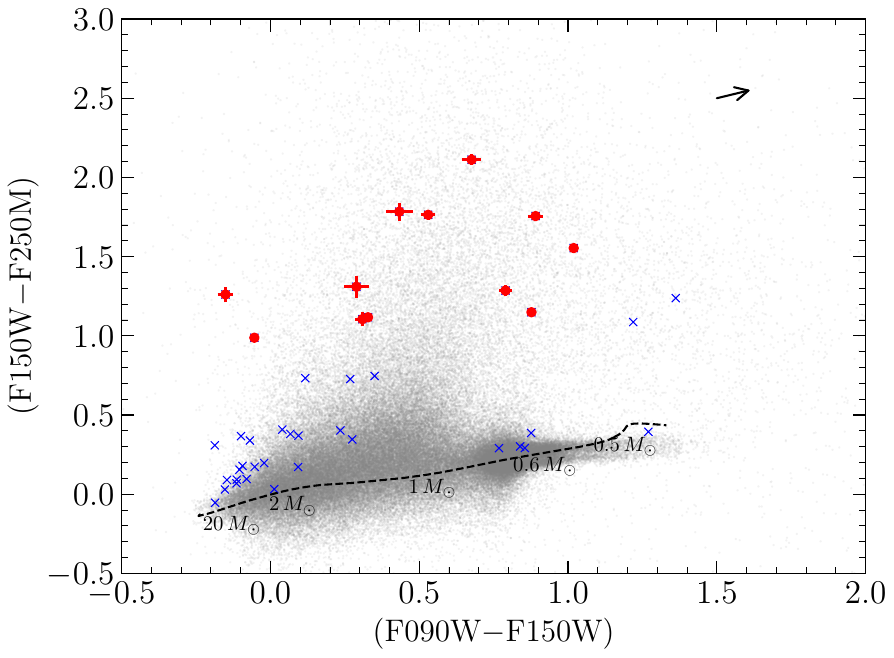}
\includegraphics[width=0.43 \textwidth]{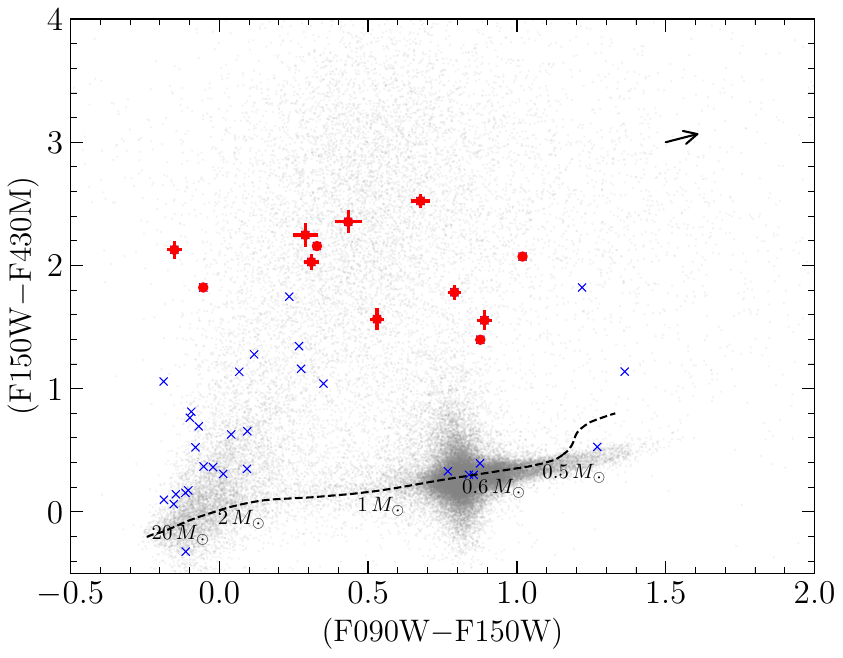}
\caption{(F090W$-$F150W) versus (F150W$-$F250M) ({\it {left}}), and (F090W$-$F150W) versus (F150W$-$F430M) ({\it {right}}) two-color diagrams of WLM. The main-sequence color locus from \cite{mist} is given by the dotted line. Blue crosses represent all sources falling within the central cluster, while red circles give the sources marked as having infrared excesses in both F250M and F430M filters, and are the pre-main sequence candidates. The stellar masses correspond to main-sequence stars at the distance and extinction adopted. Photometric uncertainties are only given for PMS candidates. }\label{ccd}
\end{figure*}

WLM is an isolated dwarf irregular galaxy in the Local Group, located 968$^{+5.4}_{-7.4}$ kpc away \citep{albers}. Its present-day metallicity is [Fe/H] $\sim$ --0.9 dex based on spectroscopy of early-type supergiants \citep{bres, urb}, and H{\scriptsize II} regions \citep{lee05}. It has an older stellar halo and a centrally concentrated younger population \citep{miniti}. Fig.\,\ref{whole} shows H$\alpha$ \citep{hunter04}, and $V$-band \citep{hunter06} images of WLM from the Lowell Observatory 1.8\,m, and 1.1\,m telescope respectively, along with a {\it Spitzer} 24\,$\mu$m image \citep{hunterir}. The H$\alpha$ emission highlights the H{\scriptsize II} regions identified in \cite{hodge95}. \cite{rubio} detected molecular clouds using Atacama Large Millimeter Array CO\,(1-0) observations (beam size$\sim$0.$\arcsec$9$\times1.\arcsec3$) near the northern H{\scriptsize II} regions.


\cite{weisz} obtained JWST images of the central region in F090W, F150W, F250M, and F430M using the Near-InfraRed Camera (NIRCam), without any dithers to conserve exposure time. Further details on the observations and photometry are in \cite{weisz} and \cite{weiszphot}, respectively. The photometric constraints from \cite{weiszphot} were applied on the F090W, F150W (and F250M/F430M when applicable) photometry. In addition to the \cite{weiszphot} cuts, a photometric uncertainty cut-off of 0.1 mag was used. However, we set the sharpness cut to be between $-$0.3 and 0.3 (see \citealt{dolphot}). Our sharpness limits are slightly relaxed, but still follow \cite{dolphot} and \cite{weiszphot} guidelines (Sec. 3.5 in that paper). We adjusted the sharpness assuming most sources are stellar within the small cluster region of interest, and any non-stellar contamination is negligible. We also visually checked the F150W images for obvious contamination.


The JWST images reveal a stellar concentration around the \cite{hodge95} H{\scriptsize II} region, [HM95]-9 (R.A.=00$^{\rm{h}}$01$^{\rm{m}}$56$^{\rm{s}}$51, Dec.=$-$15$\degr26\arcmin52.\arcsec$2). This region has a surface density of $\sim$5 sources\,pc$^{-2}$ within a 1.1$\arcsec$ radius, compared to a median of 2 sources\,pc$^{-2}$ in the surrounding field. The region within the 1.1$\arcsec$ radius from the center of [HM95]-9 is the focus of our study, and is shown in Fig.\,\ref{cl}. The reported nebular flux and line ratios of [HM95]-9 from \cite{hodge95, lee05} correspond approximately to the emitted photon flux and temperature of late O-- early B-type stars \citep{crowther}. F430M images from the surrounding region (within 2$\arcsec$, $\sim$10\,pc at the assumed distance) reveal extended diffuse emission around the cluster, likely from irradiated dust at ~700\,K. The stellar clustering is visible in the F150W images.


Fig.\,\ref{cmd} shows the color-magnitude diagram (CMD) of WLM. Highlighted are stars falling within the cluster. Stellar isochrones from \cite{mist} at [Fe/H]=$-0.9$ in the JWST filters are overlaid. The brightest sources correspond approximately to early B-type stars, which have masses around 15–18\,M$_{\odot}$. The foreground galactic extinction is adopted from \cite{mcquinn} to be {\it A$_V$}=0.1, and the extinction around the H{\scriptsize II} region is adopted to be {\it E(B-V)}=0.1 from \cite{hodge95}. The CMD of WLM shows a significant older population with a red clump. The cluster shows a younger population, with a clear upper main sequence (UMS) of massive stars ($>8$\,M$_{\odot}$), with (F090W$-$F150W)$<0$, and F090W$<23$\,mag. The mean density of UMS stars on the CMD in the cluster is more than two orders of magnitude greater than the field. The well-populated red clump seen in other regions is absent, and there are many stars located at the bottom right of the main sequence. The few central cluster stars located in the red clump and red-giant branch part of the CMD are likely background contaminants based on a qualitative and quantitative analysis of surrounding similarly sized control fields. We posit that these must by PMS stars belonging to the cluster. The presence of massive stars on the main sequence (having turn off ages less than 10\,Myr), and also nearby CO clouds strengthens our hypothesis. These  faint, red stars form the low-mass counterparts to the UMS stars, and are approaching the main sequence within the Kelvin-Helmholtz contraction timescale. At the distance to the WLM, fiducial low-mass PMS stars would potentially exist at F090W$>$26\,mag, red-ward of the main sequence locus, but slightly brighter than their main-sequence counterparts. As these stars are visible in the CMD shown in Fig.\,\ref{cmd}, we can separate them from the main sequence population based on their infrared excess from JWST data, and do so in Section 3.

\section{Pre-main sequence candidates}

Pre-main sequence stars are young stars ($<$10\,Myr) that have not yet ignited nuclear fusion. They accrete mass from a cirumcstellar disk while contracting onto the main sequence. Therefore, they can be identified based on excess emission at infrared wavelengths ($>$2\,$\mu$m; see \citealt{kenyon}) due to the disk, but also broad emission lines such as H$\alpha$, Br$\gamma$ arising from the accretion process (e.g.,  \citealt{30dor, kalariha}). Using the available JWST imaging, we can identify stars with infrared excesses by comparing them to the stellar main-sequence locus in colors containing one band redward of 2\,$\mu$m, where typically disk excesses are detected \citep{hartmann}. We follow a similar method to identify PMS candidates as \citet{jones346} here, but we use the medium-band filters F250M and F430M to represent the wide-band filters F200W and F444W used in that work. 

In Fig.\,\ref{ccd}, we plot JWST two-color diagrams. The color (F090W$-$F150W) is a proxy for the stellar spectral type, while the (F150W$-$F250M) and (F150W$-$F430M) colors can reveal excess emission at wavelengths longward of 2\,$\mu$m with respect to a pure stellar photosphere. The medium-band filter F250M is better suited for identifying infrared excesses compared to the F200W filter due to its narrower band-width and redder central wavelength covering only the region $>2.4\,\mu$m, where circumstellar disk excess is more prominent than at the central wavelength of the F200W filter (2\,$\mu$m). The F430M filter also includes the CO$_2$ absorption feature at 4.3\,$\mu$m. In Fig.\,\ref{ccd}, the UMS is clearly visible near (0,0). Evolved red stars (red supergiants, giant branch stars) from WLM are visible to the bottom right of each diagram, around (F090W$-$F150W)=0.8. 

To identify potential pre-main sequence stars, we select stars showing colors greater than 0.8 from the main-sequence locus in both (F150W$-$F250M) and (F150W$-$F430M) colors. The 0.8 value corresponds to the 3$\sigma$ limit propagated from the photometric uncertainty criteria combining three filters ($\sim$0.2\,mag). In total, 12 sources meet the excess criteria in both (F150W$-$F250M) and (F150W$-$F430M). They are tabulated in Table\,1. In the CMD (Fig.\,\ref{cmd}), most of the sources lie redward of the main sequence, with F090W magnitudes ranging from 25 to 28\,mag, consistent with objects that form main-sequence stars between 1.25--5\,M$_{\odot}$. Note that near-infrared colors, especially longward of H-band, are not commonly used to estimate stellar properties such as ages from evolutionary models due to the possible presence of circumstellar material, but also the models themselves become degenerate at lower masses \citep{belliso}. The ages based on the models of \cite{mist} are between 0.1 and 10 Myr. There also may be contaminants (e.g., reddened late type stars), or sources of error (e.g., poor photometric precision in F430M, CO$_2$ absorption) in our field, but the location of the stars in the cluster, and the significant excesses present in both F430M and F250M suggest that they are excellent PMS candidates. Plotting the positions of these stars on the F150W image (Fig.\,\ref{cl}), we see that they are located in the periphery of the cluster, towards the north. Our selection criteria should be augmented in the future by narrow-band Pa$\alpha$ photometry, which traces on-going accretion in PMS candidates.


\section{Discussion and future work} 

In this letter, we present evidence for an active star-forming cluster subtending 10\,pc in the main body of WLM. The cluster contains late O-- early-B type stars on the main sequence. H$\alpha$ emission on and around the cluster is visible. It is located near molecular clouds, and heated circumstellar dust. We hypothesize that the cluster studied here is part of the larger H{\scriptsize II} region, and akin to star-forming regions seen in the Milky Way and the Magellanic Clouds. It is ideal for follow-up with deeper and multi-band imaging (including narrow-band indicators of accretion such as Pa$\alpha$) with JWST to identify PMS stars throughout the H{\scriptsize II} region.

The cluster contains at least 12 PMS candidates presented in Table\,1. To date, the most metal-poor PMS stars known are at present-day metallicities around $-$0.45\,dex in the Milky Way star forming region, Sh 2-284 \citep{dol25}, and the Small Magellanic Cloud \citep{jones346, demarchi346, demarchi602}. The latter are also the most distant ($\sim$61\,kpc). The discovery of these candidates extends this by a factor of two in metallicity, and more than a factor of 10 in distance, and suggests PMS candidates can be detected with JWST in local group galaxies up to distances of 1\,Mpc. This demonstrates that PMS candidates with disks akin to those found in the Milky Way and Magellanic Clouds are present in more metal-poor star-forming regions, and can be detected at such distances using current observational facilities. Based on the turn-off age of early B-type stars, and isochrone-based PMS star ages, we suggest that the cluster and PMS candidates are less than 10\,Myr old. 
With the limitations of the current data, the density of PMS stars in the cluster appears to resemble that of star-forming regions in the Magellanic Bridge at around 1/5\,$Z_{\odot}$ \citep{bridge} and low-density star forming regions in the solar neighborhood \citep{heidermann}. Although individual PMS candidates may not be well resolved with the current angular resolution (ranging between 30000 and 80000 AU in the F090W and F250M filters respectively\footnote{https://jwst-docs.stsci.edu/jwst-near-infrared-camera/nircam-performance/nircam-point-spread-functions\#NIRCamPointSpreadFunctions-PSFFWHM}), we note that very few young stellar objects were found with companions between 2000-40000 AU in the Magellanic Clouds by \cite{kalarizorro}, suggesting that contamination from neighboring sources may not be a significant issue. We hypothesize that further on-going star formation in the surrounding region could be detected with deeper multi-band JWST imaging with higher completeness fractions and, particularly, with the addition of a Pa$\alpha$ filter to identify on-going accretion. If so, star formation in those regions might represent a second more distributed generation, following the clustered mode of star formation detected here.

\begin{deluxetable*}{llllllll}
\tablenum{1}
\tablecaption{PMS candidates identified in WLM}
\tablewidth{0pt}
\tablehead{
\colhead{No.} &
  \colhead{ID$^1$} &
  \colhead{R.A.} &
  \colhead{Dec.} & \colhead{F090W} &\colhead{F150W} &\colhead{F250M} & \colhead{F430M}  \\
    \colhead{ } &
  \colhead{} &
  \colhead{} & \colhead{} & \colhead{(mag)} &\colhead{(mag)} &\colhead{(mag)} & \colhead{(mag)} 
}
\startdata
  9 & 172025 & 00:01:56.49 & $-$15:26:51.4 & 26.968$\pm$0.014 & 26.659$\pm$0.02 & 25.554$\pm$0.041 & 24.631$\pm$0.061\\
  3 & 59473 & 00:01:56.51 & $-$15:26:51.8 & 25.421$\pm$0.005 & 25.475$\pm$0.009 & 24.487$\pm$0.021 & 23.654$\pm$0.03\\
  1 & 51486 & 00:01:56.52 & $-$15:26:52.1 & 25.257$\pm$0.005 & 24.929$\pm$0.006 & 23.812$\pm$0.014 & 22.772$\pm$0.016\\
  2 & 56000 & 00:01:56.53 & $-$15:26:52.1 & 25.69$\pm$0.006 & 24.813$\pm$0.006 & 23.664$\pm$0.011 & 23.416$\pm$0.026\\
  13& 281242 & 00:01:56.53 & $-$15:26:52.4 & 27.509$\pm$0.022 & 27.22$\pm$0.035 & 25.91$\pm$0.061 & 24.973$\pm$0.092\\
  8 & 162894 & 00:01:56.53 & $-$15:26:53.0 & 27.052$\pm$0.015 & 26.522$\pm$0.019 & 24.758$\pm$0.023 & 24.958$\pm$0.085\\
  12 & 196063 & 00:01:56.54 & $-$15:26:52.6 & 27.45$\pm$0.02 & 26.774$\pm$0.024 & 24.662$\pm$0.023 & 24.251$\pm$0.048\\
  6 & 135739 & 00:01:56.54 & $-$15:26:52.4 & 26.632$\pm$0.011 & 26.783$\pm$0.024 & 25.522$\pm$0.044 & 24.656$\pm$0.068\\
  10 & 175422 & 00:01:56.54 & $-$15:26:53.0 & 27.352$\pm$0.019 & 26.461$\pm$0.018 & 24.705$\pm$0.022 & 24.906$\pm$0.081\\
  5 & 97479 & 00:01:56.54 & $-$15:26:52.0 & 26.699$\pm$0.012 & 25.68$\pm$0.01 & 24.126 $\pm$0.015 & 23.608$\pm$0.027\\
  7 & 154781 & 00:01:56.55 & $-$15:26:52.3 & 27.113$\pm$0.016 & 26.323$\pm$0.016 & 25.037$\pm$ 0.031 & 24.541$\pm$0.059\\
  14 & 340213 & 00:01:56.56 & $-$15:26:52.3 & 27.793$\pm$0.026 & 27.36$\pm$0.037 & 25.576$\pm$0.042 & 25.006$\pm$0.086\\
\enddata
\tablecomments{$^1$ From \cite{weiszphot}. }
\end{deluxetable*}


\begin{acknowledgments}

We thank the referee for a constructive report, which greatly helped improve this work. This work is based on observations made with the NASA/ESA/CSA James Webb Space Telescope. The data were obtained from the Mikulski Archive for Space Telescopes at the Space Telescope Science Institute, which is operated by the Association of Universities for Research in Astronomy, Inc., under NASA contract NAS 5-03127 for JWST. The JWST data presented in this article were obtained from the Mikulski Archive for Space Telescopes (MAST) at the
Space Telescope Science Institute. The specific observations analyzed can be accessed via \dataset[doi: 10.17909/cn6n-xg90].
The work of V.M.K. is supported by NOIRLab, which is managed by the Association of Universities for Research in Astronomy (AURA) under a cooperative agreement with the U.S. National Science Foundation. V.M.K. thanks Deidre Hunter for comments. M.R. wishes to acknowledge partial support from ANID(CHILE) through Basal FB210003. JSV acknowledges support from grant ST/V000233/1 (PI: Vink).
\begin{figure}
    \centering
    \includegraphics[width=0.5\linewidth]{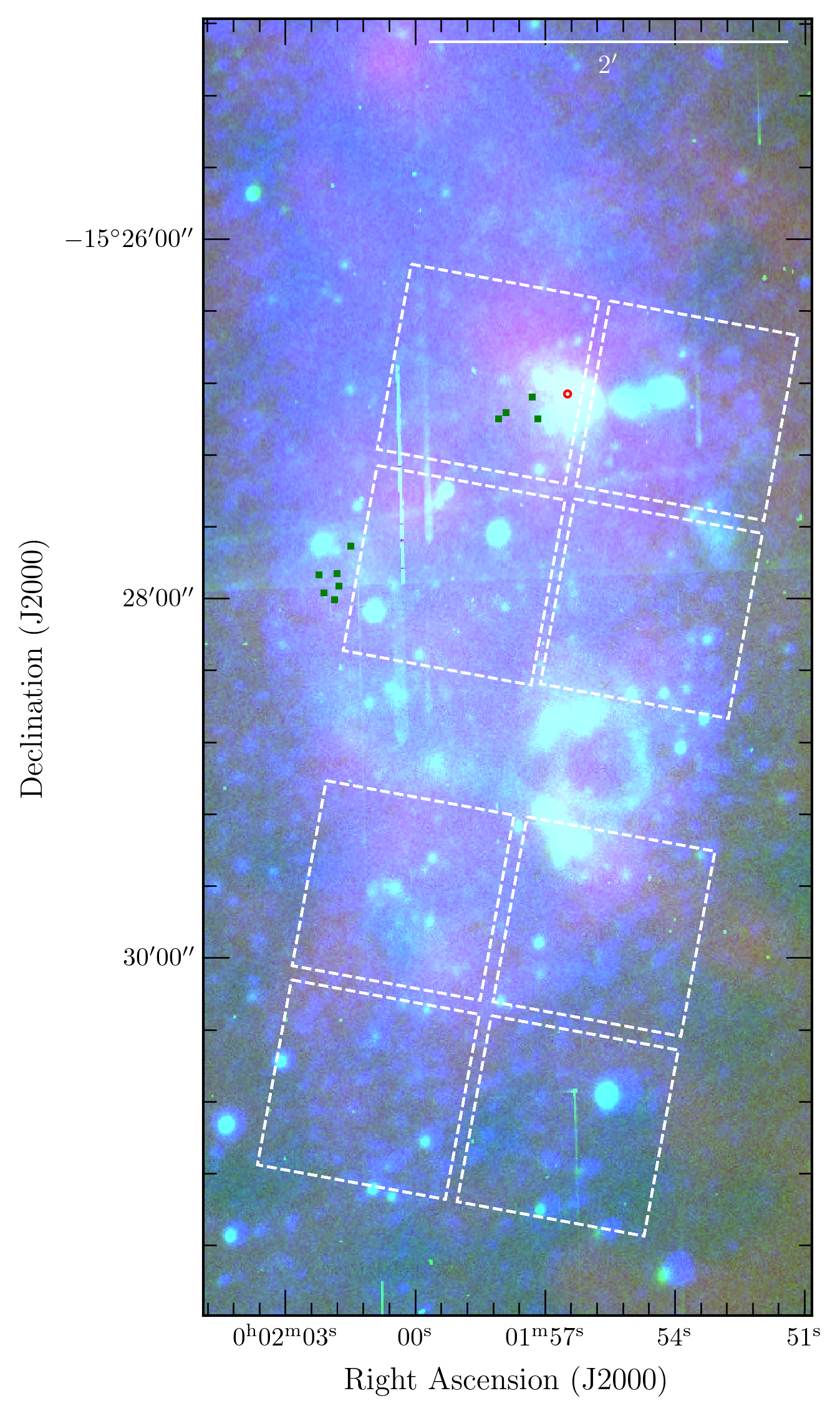}
    \caption{Enter Caption}
    \label{fig:enter-label}
\end{figure}

\end{acknowledgments}

%

\vspace{5mm}
\facilities{JWST}





\bibliography{sample631}{}
\bibliographystyle{aasjournal}



\end{document}